\documentstyle[prb,aps,epsfig]{revtex}

\begin{document}
\twocolumn[\hsize\textwidth\columnwidth\hsize\csname 
@twocolumnfalse\endcsname 

\title{Electron localization in the insulating state: \\ Application to
crystalline semiconductors}

\author{Claudia Sgiarovello, Maria Peressi, and Raffaele Resta}

\address{Istituto Nazionale di Fisica della Materia (INFM) and Dipartimento
di Fisica Teorica, Universit\`a di Trieste, Strada Costiera 11, I--34014
Trieste, Italy\cite{nota1} \\ and Institut Romand de Recherche Num\'erique en
Physique des Mat\'eriaux (IRRMA), CH--1015 Lausanne,
Switzerland\cite{nota2}}

\maketitle

\begin{abstract} We measure electron localization in different materials by
means of a ``localization tensor'', based on Berry phases and related
quantities. We analyze its properties, and we actually compute such tensor
from first principles for several tetrahedrally coordinated semiconductors. 
We discuss the trends in our calculated quantity, and we relate our findings
to recent work by other authors. We also address the ``hermaphrodite
orbitals'', which are localized (Wannier--like) in a given direction, and
delocalized (Bloch--like) in the two orthogonal directions: our tensor is
related to the optimal localization of these orbitals.  We also prove
numerically that the decay of the optimally localized hermaphrodite
orbitals is exponential.  \end{abstract}

]

%\newpage

\section{Introduction} \label{sect:intro}

A nonmetal is distinguished from a metal by its vanishing conductivity at low
temperature and low frequency: we use here the term ``insulator'' to include
any nonmetal, like the semiconducting materials which are the case studies
actually addressed in this work.

Within classical physics, the qualitative difference between an insulator and
a metal is attributed to the nature of the electronic charge: either
``bound'' (Lorentz model for insulators) or ``free'' (Drude model for
metals). In other words, electrons are {\it localized} in insulators and {\it
delocalized} in metals.  In a milestone paper published in 1964, W. Kohn
characterized the insulating state of matter in a way which is reminiscent of
the classical picture: he gave evidence that the main feature determining the
insulating behavior of matter is electron localization in the ground--state
wavefunction.\cite{Kohn} Although this work mainly addressed {\it
correlated} many--electron systems, its message is very relevant even for
materials where an independent--electron description is quite adequate, as the
semiconductor crystals studied here. Recently a novel measure of
electron localization---different from Kohn's one---was proposed by Resta and
Sorella,\cite{rap107} hereafter cited as RS. Their approach is deeply rooted
into the modern theory of
polarization.\cite{modern,rap_a12,Ortiz94,rap_a18,position}

Metals and insulators reveal their qualitative difference when static
dielectric polarization is addressed. Suppose we expose a finite macroscopic
sample to an electric field, say inserting it in a charged capacitor. In
metals polarization is trivial: universal, material--independent, due to
surface phenomena only (screening by free carriers). Therefore polarization
in metals is {\it not} a bulk phenomenon.  The opposite is true for
insulators: macroscopic polarization is a nontrivial, material--dependent,
bulk phenomenon. We can therefore phenomenologically characterize an
insulator, in very general terms, as a material whose ground wavefunction
sustains a nonzero bulk macroscopic polarization whenever the electronic
Hamiltonian is non centrosymmetric. If the Hamiltonian is instead
centrosymmetric, the polarization vanishes but remains a well defined bulk
property, at variance with the metallic case. The phenomenological link
between macroscopic polarization and insulating behavior was first pointed
out and exploited---taking advantage of the modern theory of polarization
\cite{modern,rap_a12,Ortiz94,rap_a18,position}---by RS in 1999. This approach 
is based on Berry phases and related concepts.\cite{rap_a20} Even the RS
paper, like Kohn's 1964 one, mostly concerns correlated systems. Furthermore,
in order to keep the presentation simple and concise, most results are
explicitly shown in one dimension, while the $d$--dimensional formulation is
only sketched in the final paragraphs of RS.  In the present paper we provide
more details on how the RS theory of localization works in three dimensions,
specializing to a system of noninteracting electrons, like the band
insulators chosen as case studies here.

Some other important papers must be mentioned at this point. In 1997 Marzari
and Vanderbilt,\cite{Marzari97} hereafter cited as MV, while not addressing
metals at all (and hence their difference from insulators), establish
nonetheless some results which are relevant to the present viewpoint. In a
very recent comprehensive paper\cite{Souza00} Souza, Wilkens, and
Martin---hereafter cited as SWM---generalize and extend in various ways the
main finding of RS: we adopt here some of their notations. Finally, after
this work was completed, we became aware of Ref.~\onlinecite{He01}, whose
conclusions bear some implications for our results shown in
Sect.~\ref{sect:herma}.

The paper is organized as follows. In Sect.~\ref{sect:many} we define the
basic ingredients providing both polarization and localization, namely the
expectation values of the many--body phase operators $z_N^{(\alpha)}$ for the
three Cartesian coordinates, Eq.~(\ref{main}). In
Sect.~\ref{sect:loca}, following RS, we show how the modulus of
$z_N^{(\alpha)}$ defines a very meaningful quantity, the
localization tensor, for which we adopt the SWM notations: such tensor is
finite in insulators and diverges in metals. In Sect.~\ref{sect:prope} we
discuss the main properties of the localization tensor, and in
Sect.~\ref{sect:calcu} we present first--principle calculations for several
elemental and binary semiconductors: the main trends are analyzed. In
Sect.~\ref{sect:herma} we calculate orbitals which are optimally localized in
a given direction, and whose average quadratic spread coincides with the
localization tensor. We also heuristically check the exponential localization
of these orbitals, which we call ``hermaphrodite orbitals''.  In
Sect.~\ref{sect:conclu} we draw our main conclusions. In the Appendix we
consider a molecule or a cluster an we discuss our localization tensor
therein, showing its relationship to some results of Boys localization
theory,\cite{Boys60} well known in quantum chemistry.\cite{Pipek89}

\section{Many--body phase operators} \label{sect:many}

We are addressing here, as it is done by MV, a crystalline system of
independent--electrons, having in mind a Kohn--Sham scheme. The properties of
interest, namely, macroscopic polarization and electron localization, are not
properties of the individual KS orbitals: instead, they are global properties
of the occupied KS manifold. As shown in Refs.~\onlinecite{rap107,position},
it proves formally convenient to deal with a many--body wavefunction $\Psi$,
obtained as a Slater determinant of occupied orbitals. This
determinant is uniquely determined by the manifold of the occupied orbitals
and is invariant by unitary transformation of these orbitals among
themselves: for instance, in insulating crystals, an important transformation
of this class converts the occupied Bloch orbitals into Wannier
functions.\cite{Blount} Quantities which can be expressed solely in terms of
$\Psi$ are invariant in form under such transformations.

Throughout this work---with the exception of the Appendix---we adopt periodic
Born--von--K\`arm\`an boundary conditions (BvK) on a large cell, multiple of
the crystalline elementary cell. The quantities of interest are intensive and
have a well defined thermodynamic limit, while the wavefunction itself
becomes an ill--defined mathematical object in that limit.  

For the sake of simplicity, we assume a simple cubic cell of side $a$ and a
large BvK cell of side $L = Ma$. More general structures can be dealt with
using scaling, similarly to what shown {\it e.g.} in
Ref.~\onlinecite{rap_a12} or in SWM.  The thermodynamic limit corresponds to
$M \rightarrow \infty$, while practical calculations are performed at finite,
and possibly large, $M$ values.  The spinorbitals $\psi$ (spin--up)  and
$\overline{\psi}$ (spin--down) may be chosen of the Bloch form.  In the
finite system there are $M^3$ allowed Bloch vectors ${\bf q}_s$, arranged on
a regular mesh in the unit reciprocal cell, where $s \equiv (s_1, s_2, s_3)$
and \begin{equation}  {\bf q}_s = \frac{2 \pi}{M a} (s_1, s_2, s_3) , \quad
s_\alpha = 0,1,\dots, M\! -\!1 . \label{points} \end{equation} We adopt a
plane--wave--like normalization for the Bloch orbitals: \begin{equation}
\langle \psi_{n{\bf q}_s} | \psi_{n'{\bf q}_{s'}} \rangle = \frac{1}{L^3}
\int_{\mbox{\scriptsize BvK cell}}  \!\!\!\!\!\!\! d {\bf r} \; \psi^*_{n{\bf
q}_s}({\bf r}) \psi_{n'{\bf q}_{s'}}({\bf r}) = \delta_{nn'} \delta_{ss'},
\end{equation} 

If the system is insulating with $n_b$ doubly occupied bands, there are $N =
2 n_b M^3$ independent spinorbitals, out of which we write a
single--determinant many--body wavefunction for $N$ electrons: 
\begin{equation} \Psi = {\sf
A} \prod_{n,s} \frac{1}{L^3} \psi_{n{\bf q}_s} \overline{\psi}_{n{\bf q}_s},
\label{slater} \end{equation} where the product runs over all occupied bands 
and all mesh
points, ${\sf A}$ is the antisymmetrizer operator, and the factor ensures
that the $N$--body wavefunction is normalized to one on the hypercube of side
$L$. If, instead, the system is metallic, then the many--body wavefunction
$\Psi$ can still be written in the form of Eq.~(\ref{slater}), but where not
all the Bloch vectors of a given band are included in the product. 

According to Refs.~\onlinecite{rap107,position}, the key quantities to deal
with both macroscopic polarization and electron localization are expectation
values of ``many--body phase operators''. For a three--dimensional system
there are three such operators, one for each Cartesian direction. We indicate
as  $z_N^{(\alpha)}$, where $\alpha$ is a Cartesian index, their
ground--state expectation values: \begin{equation} z_N^{(x)} = \langle \Psi | 
{\rm
e}^{i\frac{2\pi}{L} \sum_{i=1}^N x_i} | \Psi \rangle , \label{main} 
\end{equation} and
analogously for $y$ and $z$ directions. This remarkably compact expression is
very general and applies as it stands even to correlated and/or disordered
systems: here we specialize to a crystalline system of independent electrons,
whose wavefunction $\Psi$ assumes the form of Eq.~(\ref{slater}), where the
product indices have to be differently specified in the insulating and
metallic cases. 

We may conveniently recast $z_N^{(x)}$ as an overlap \begin{equation}
{z}_N^{(x)} = \langle \Psi | \tilde{\Psi} \rangle , \label{main2}
\end{equation} where $\tilde{\Psi}$ is the Slater determinant of a different
set of Bloch spinorbitals: \begin{equation} \tilde{\psi}_{n{\bf q}_s}({\bf
r}) = {\rm e}^{i\frac{2\pi}{Ma}x} \psi_{n{\bf q}_s}({\bf r}) , \end{equation}
and analogously for the bar (spin--down) ones.  According to a well known
theorem, the overlap between two single--determinant wavefunctions is equal
to the determinant of the $N \times N$ overlap matrix built out of the
occupied spinorbitals. Since the overlaps between different--spin
spinorbitals vanish, and those between equal--spin ones are identical in
pairs, we can write \begin{equation} {z}_N^{(x)} = ( \mbox{det} \;
{\cal S} )^2, \label{det} \end{equation} where ${\cal S}$ is the overlap
matrix between spatial orbitals, having size $ N/2 = n_b M^3$.  Its elements
are: \begin{eqnarray} {\cal S}_{n{\bf q}_s,n'{\bf q}_{s'}} = \frac{1}{L^3}
\int_{\mbox{\scriptsize BvK cell}}  \!\!\!\!\!\!\! d{\bf r} \; \psi^*_{n{\bf
q}_s}({\bf r}) \tilde{\psi}_{n'{\bf q}_{s'}}({\bf r}) \nonumber \\ =
\frac{1}{L^3} \int_{\mbox{\scriptsize BvK cell}}  \!\!\!\!\!\!\! d{\bf r} \;
u^*_{n{\bf q}_s}({\bf r}) u_{n'{\bf q}_{s'}}({\bf r}) {\rm e}^{i
(\frac{2\pi}{Ma}x + {\bf q}_{s'} \cdot {\bf r} - {\bf q}_s \cdot {\bf r} )},
\label{overlap} \end{eqnarray} where the $u$'s are the periodic functions in
the Bloch orbitals.

The matrix ${\cal S}$ is very sparse: in fact, given the geometry of the
${\bf q}_s$'s on the regular reciprocal mesh
(see Eq.~(\ref{points}), the overlap integrals in
Eq.~(\ref{overlap}) are nonvanishing only if $s_1 = s_1' + 1$, $s_2 = s_2'$, 
and $s_3 = s_3'$. We express the nonvanishing elements in terms of a small
overlap matrix $S$, of size $n_b \times n_b$: \begin{equation} S_{nn'}({\bf
q,q'}) = \langle u_{n{\bf q}} | u_{n'{\bf q'}} \rangle =
\frac{1}{a^3}\int_{\mbox{\scriptsize cell}} \!\!\! d{\bf r} \; u^*_{n{\bf
q}}({\bf r}) u_{n'{\bf q}'}({\bf r}) . \label{ovp2} \end{equation}  Owing to
the sparseness of ${\cal S}$, its determinant factors into products of
determinants of small matrices $S$.  

In the insulating case we use the wavefunction of Eq.~(\ref{slater}), 
where all the
Bloch vectors of a given band are occupied: the factorization is then 
\begin{equation}
{z}_N^{(x) 1/2} = \mbox{det} \; {\cal S} = \prod_{s} \mbox{det} \;
S({\bf q}_{s_1+1,s_2,s_3}, {\bf q}_{s_1,s_2,s_3}) \label{facto} . 
\end{equation} 
We get a
more compact notation upon defining \begin{equation} \Delta {\bf q}^{(x)} = 
\frac{2 \pi}{L}
(1,0,0) = \frac{2 \pi}{M a} (1,0,0) , \label{dq} \end{equation} which is the 
vector
connecting nearest neighbor {\bf q} points in the $x$ direction. We have
then: \begin{equation} {z}_N^{(x) 1/2} = \prod_{s} \mbox{det} \; S({\bf
q}_s\!+\!\Delta {\bf q}^{(x)}, {\bf q}_s) \label{facto1} ; \end{equation} 
In the metallic
case, instead, the $z_N^{(\alpha)}$'s are identically zero.  This is easily
understood by looking at the simple case with only one band.  Suppressing the
band index the small overlap matrix becomes a c--number $S({\bf q,q'})$, and
Eq.~(\ref{facto}) becomes a product of c--numbers, with no determinant to 
evaluate.
In an insulator this product runs over the whole ${\bf q}_s$ mesh, and all
factors are nonzero; in a metal the analogous product runs only on the ${\bf
q}_s$'s within the Fermi surface. Looking at the definition of $S({\bf
q,q'})$, Eqs.~(\ref{overlap}) and (\ref{ovp2}), it is clear that there exists 
at least one
occupied ${\bf q}_s$, adjacent to the Fermi surface, such that $S({\bf
q}_s,{\bf q}_{s'})$ vanishes for all {\it occupied} $s'$. This is enough to
imply that $z_N^{(\alpha)}$ vanishes as well.

The complex numbers $z_N^{(\alpha)}$ are ground--state expectation values,
and do not access any spectral information. Yet they qualitatively
discriminate between insulators and metals: they are in fact nonvanishing in
the former materials, and vanishing in the latter ones. This shows, according
to RS, that there is a qualitative difference in the organization of the
electrons in their ground state. It is remarkable that, in the present case,
such difference shows up already at {\it finite} $N$, before the
thermodynamic limit is taken.

\section{Electron localization} \label{sect:loca}

In centrosymmetric materials the expectation values $z_N^{(\alpha)}$ are real
(provided the origin is chosen at a centrosymmetric site), while in
noncentrosymmetric materials they are in general complex: their phases define
then the Cartesian components of the macroscopic polarization in suitable
units.\cite{rap107,position}
In the metallic case the $z_N^{(\alpha)}$'s vanish and the
polarization is ill defined, in agreement with the phenomenological viewpoint
illustrated in Sect.~\ref{sect:intro}. We address electron localization using
the moduli of these same $z_N^{(\alpha)}$'s. Following RS,
electron localization is measured by a squared localization length in one
dimension, and by a ``localization tensor'' in three dimensions. This tensor
is an intensive quantity, has the dimensions of a squared length, and
measures the localization of the many--electron system as a whole: in the
present case, it is a global property of the occupied KS manifold. The
localization tensor is finite for insulators and diverges for metals.  

In the very recent SWM paper\cite{Souza00} it is shown, among other things,
that the RS localization tensor is related to the mean--square quantum
fluctuation of the polarization: it is a second cumulant moment, which can be
very elegantly extracted from a moment generating function. We adopt
throughout notations inspired by SWM, and we indicate the localization tensor
as $\langle r_\alpha r_\beta \rangle_{\rm c}$, where the subscript stays for
``cumulant''. For a material having cubic or tetrahedral symmetry, like the
semiconductors considered in the present case studies, the localization
tensor is isotropic: its only independent element is $\langle x^2
\rangle_{\rm c}$. Its expression is provided by RS, whose Eq.~(18) we recast
here as \begin{equation} \langle x^2 \rangle_{\rm c} = - \frac{1}{N}
\left(\frac{L}{2\pi}\right)^2 \ln |{z}_N^{(x)}|^2 , \label{guess}
\end{equation} and the thermodynamic limit is understood. For a metal
$z_N^{(x)}$ vanishes and the localization tensor is formally infinite, even
at finite $N$. For an insulator, whose wavefunction has the form of
Eq.~(\ref{slater}), we get from Eq.~(\ref{facto1}) \begin{equation}
|{z}_N^{(x)}| = \prod_{s} \mbox{det} \; S^\dagger ({\bf q}_s, {\bf
q}_s\!+\!\Delta {\bf q}^{(x)}) S({\bf q}_s, {\bf q}_s\!+\!\Delta {\bf
q}^{(x)}) \label{facto3} ;  \end{equation} \begin{equation} \langle x^2
\rangle_{\rm c} =  - \left( \frac{a}{2 \pi} \right)^2 \frac{1}{2 n_b M} \ln 
|{z}_N^{(x)}|^2  . \label{mdiago} \end{equation}

Eqs.~(\ref{facto3}) and (\ref{mdiago}) are the typical expressions 
implemented in our
test--case calculations discussed below. The thermodynamic limit is 
obtained as
usual for $M \rightarrow \infty$ and takes, not surprisingly, the form of an
integral performed over the reciprocal unit cell, or equivalently over the
first Brillouin zone. The integral is: \begin{eqnarray} \langle x^2
\rangle_{\rm c} = \frac{a^3}{n_b (2\pi)^3} \int d{\bf q} \; \left( \sum_n
\langle \frac{\partial}{\partial q_x} u_{n{\bf q}} | \frac{\partial}{\partial
q_x} u_{n{\bf q}} \rangle \right. \nonumber \\ \left.  - \sum_{n,n'} \langle
u_{n{\bf q}} | \frac{\partial}{\partial q_x} u_{n'{\bf q}} \rangle \langle
\frac{\partial}{\partial q_x} u_{n'{\bf q}} | u_{n{\bf q}} \rangle \right) . 
\label{mom1} \end{eqnarray} The proof is relatively straightforward, starting
from Eq.~(\ref{mom1}) and discretizing integrals and derivatives on the mesh
defined in Eq.~(\ref{points}).

Expressions such as Eq.~(\ref{mom1}) and similar ones had appeared in the
literature before,\cite{Blount} in relationship to Wannier functions.  By
means of an expression of this kind, MV define a ground--state quantity
$\Omega_{\rm I}$ which sets a lower bound for the second (spherical) moments
of the Wannier functions.\cite{Marzari97} More precisely, for an insulator
with $n_b$ occupied bands (hence $n_b$ Wannier functions per cell) such
second moment is no smaller in average than $\Omega_{\rm I}/n_b$. It is worth
mentioning at this point that the logic of the MV paper goes backwards with
respect to the present approach: first they provide a continuum theory, and
then they discretize for computational purposes. Their discretization is
different from Eq.~(\ref{mdiago}), which emerges naturally from the present
formulation starting from the remarkably compact Eq.~(\ref{guess}). Both
discretizations obviously converge to the same $M \rightarrow \infty$ limit:
their convergence properties are different, though.

Specializing MV to a cubic material, RS have found the simple relationship
$\Omega_{\rm I} = 3 n_b \langle x^2 \rangle_{\rm c}$: notice that $\langle
x^2 \rangle_{\rm c}$ is intensive, while $\Omega_{\rm I}$ is not such. 
Building upon MV's work, we are now ready to generalize the localization
tensor to materials of arbitrary symmetry: \begin{eqnarray} \langle r_\alpha
r_\beta \rangle_{\rm c} = \frac{V_c}{n_b (2\pi)^3} \int d{\bf q} \;
\left( \sum_n \langle \frac{\partial}{\partial q_\alpha} u_{n{\bf q}} |
\frac{\partial}{\partial q_\beta} u_{n{\bf q}} \rangle \right. \nonumber \\
\left.  - \sum_{n,n'} \langle u_{n{\bf q}} | \frac{\partial}{\partial
q_\alpha} u_{n'{\bf q}} \rangle \langle \frac{\partial}{\partial q_\beta}
u_{n'{\bf q}} | u_{n{\bf q}} \rangle \right) , \label{mom2} \end{eqnarray}
where $V_c$ is the cell volume.  Notice that the imaginary part of the
integrand in Eq.~(\ref{mom2}), being antisymmetric in ${\bf q}$, cancels in 
the integral, such that the localization tensor is real. Even the offdiagonal
elements, as defined in Eq.~(\ref{mom2}), have a finite--$N$ counterpart in
terms of many--body phase operators.

For an insulating crystal of arbitrary symmetry, $\Omega_{\rm I}$ as defined
by MV equals $n_b$ times the trace of our localization tensor $\langle
r_\alpha r_\beta \rangle_{\rm c}$. In a metal, expressions like
Eqs.~(\ref{mom1}) and (\ref{mom2}) do not make much sense, consistently with 
the fact that our
finite--$N$ expression, Eq.~(\ref{guess}), is formally infinite at any $N$ 
value.

\section{Properties of the localization tensor} \label{sect:prope}

We have already emphasized that the localization tensor is a property of the
occupied KS manifold as a whole. The main quantity which defines such
manifold is the (spin--integrated) single--particle density matrix $\rho$,
which coincides with twice the projector $P$ over the occupied KS orbitals: 
this projector is invariant by unitary transformations of the orbitals. Using
Bloch eigenfunctions the projector reads, for an insulator with $n_b$
occupied bands: \begin{equation} P({\bf r,r'}) = \frac{1}{2} \rho({\bf r,r'}) =
\frac{1}{(2\pi)^3} \sum_{n=1}^{n_b} \int d{\bf q} \; \psi_{n{\bf q}} ({\bf r})
\psi_{n{\bf q}}^* ({\bf r'}) . \label{proj} \end{equation} The localization 
tensor (in
the thermodynamic limit) has been written as a Brillouin--zone integral in
Eq.~(\ref{mom2}). This integral can be identically transformed into a 
particularly
simple expression whose only ingredient is $P$: \begin{equation} \langle 
r_\alpha r_\beta
\rangle_{\rm c} = \frac{1}{2n_b} \int_{\mbox{\scriptsize cell}} \!\!\! d {\bf
r} \int_{\mbox{\scriptsize all space}} \!\!\!\!\!\!\!\!\!\!\!\! d {\bf r}' \;
({\bf r - r'})_\alpha ({\bf r - r'})_\beta \, |P({\bf r}, {\bf r}')|^2
\label{mproj} , \end{equation} which is the second moment of the (squared) 
density matrix
in the coordinate ${\bf r - r'}$. The proof of the equivalence between
Eq.~(\ref{mproj}) and Eq.~(\ref{mom2}) can be worked out using the same 
algebra appearing
in Ref.~\onlinecite{Blount}: for a different argument proving the same
result, see the Appendix.

We have arrived at Eq.~(\ref{mproj}) considering an insulating crystal so far. 
In this case we know, under general
arguments,\cite{Kohn59,desCloizeaux64,Kohn96,Ismail99} that $P({\bf r,r'})$
is asymptotically exponential in the argument $|{\bf r - r'}|$: this confirms
that the integral over all space  in Eq.~(\ref{mproj}) converges and the
localization tensor is therefore finite. At this point, it is worthwhile to
apply the general form of Eq.~(\ref{mproj}) to the metallic case.  For the 
simplest
metal of all, the free electron gas, the density matrix is known
exactly:\cite{JonesMarch} \begin{equation} P({\bf r,r'}) = \frac{1}{2} 
\rho({\bf r,r'})
=\frac{3 n_0}{2} \frac{j_1(k_{\rm F} |{\bf r-r'}|)}{ k_{\rm F} |{\bf r-r'}|}
.  \label{metal} \end{equation} Replacement of Eq.~(\ref{metal}) into
Eq.~(\ref{mproj}) 
results in a
diverging integral, thus confirming that our localization tensor is formally
infinite in this paradigmatic metal. Other, more realistic, metals feature
this same divergence.

The fact that the density matrix $\rho({\bf r,r'})$ is short--range in the
variable ${\bf r - r'}$ has been named ``nearsightedness'' by W. 
Kohn.\cite{Kohn96} The second moment expression in Eq.~(\ref{mproj}) shows
that our localization tensor is indeed a meaningful quantitative measure of
such nearsightedness. We are going to analyze below the major trends over an
important class of materials: tetrahedral semiconductors. We mention at this
point that a conceptually different measure of the nearsightedness of a given
electronic ground state focuses instead on the exponent governing the
exponential decay of $\rho({\bf r,r'})$ in insulators: some case studies have
been recently investigated.\cite{Ismail99,Stephan00}

We have already observed that some of our findings are closely related to the
previous work by MV. These authors' main interest were the ``optimally
localized Wannier functions'', {\it i.e.} those localized orbitals which
minimize the average spherical moment. They prove, among other things, that
such moment is strictly larger than the trace of our localization tensor.
Building on their results, it is straightforward to attribute a similar
meaning to the tensor itself: for any transformation of the occupied orbitals
into a set of unitarily equivalent ones, the second moment in a given
direction can be no smaller than the localization tensor, projected on that
direction.  

Since we are going to apply our results to cubic materials only, we focus on
those orbitals which minimize in average the quadratic spread (second moment)
in the $x$ coordinate. The present formalism makes the definition of these
orbitals particularly simple: they are in fact the eigenfunctions of the
position operator $x$, projected over the occupied manifold. Calling $\Xi =
PxP$ this operator, its expression in the Schr\"odinger representation is:
\begin{equation} \Xi({\bf r, r'}) = \int_{\mbox{\scriptsize all space}}
\!\!\!\!\!\!\!\!\!\!\!\! d{\bf r''} \; P({\bf r, r''}) x'' P({\bf r'', r'}) .
\label{proj2} \end{equation} Notice that $x$ is {\it incompatible} with BvK
boundary conditions and its matrix elements over Bloch states are ill
defined; nonetheless, $\Xi$ is---in insulators---a well defined operator,
which maps any vector of the occupied manifold into another vector of the
same manifold.  This fact owes to the exponential localization of $P$ in
Eq.~(\ref{proj2}).

The relationship between $\Xi$ and the orbitals optimally localized in the
$x$ direction is easily proved borrowing some results from MV; for a
different argument leading to the same proof, see the Appendix. We also
notice an important difference with respect to the three--dimensional
localization explicitly considered by MV. While the trace of the localization
tensor provides a {\it lower bound} for three--dimensional localization, its
element $\langle x^2 \rangle_{\rm c}$ provides instead a genuine {\it
minimum} for one--dimensional localization (in a cubic material). This
qualitative difference owes to the fact that, while one can manifestly
diagonalize $PxP$, one cannot simultaneously diagonalize $PxP$, $PyP$, and
$PzP$.

We end this Section about the properties of the localization tensor with a
most important issue: is $\langle r_\alpha r_\beta \rangle_{\rm c}$ a
measurable quantity? The answer, due to SWM, is ``yes''. They prove the
identity: \begin{equation} \langle r_\alpha r_\beta \rangle_{\rm c} = 
\frac{\hbar V_c}{2
\pi e^2 n_b} \int_0^{\infty} \frac{d \omega}{\omega} \, {\rm Re} \;
\sigma_{\alpha\beta}(\omega) , \label{swm} \end{equation} where 
$\sigma_{\alpha\beta}$ is
the conductivity tensor. Notice that the left hand side, as emphasized
throughout the present work, is a property of the electronic {\it ground
state}, while the right hand side is a measurable property related to
electronic {\it excitations}: therefore Eq.~(\ref{swm}) must be regarded as a 
sum rule.  The frequency integral in Eq.~(\ref{swm}) diverges in metals 
and is finite
in insulators, as obviously expected. Since in the latter materials there is
a gap for electronic excitations, Eq.~(\ref{swm}) immediately leads to the
inequality: \begin{equation} \langle r_\alpha r_\beta \rangle_{\rm c} < 
\frac{\hbar V_c}{2
\pi e^2 n_b \varepsilon_g } \int_0^{\infty} d \omega \, {\rm Re} \;
\sigma_{\alpha\beta}(\omega) , \label{swm2} \end{equation} where 
$\varepsilon_g$ is the
direct gap. Using then the oscillator--strength sum rule, Eq.~(\ref{swm2}) 
for a
cubic material is cast as: \begin{equation} \langle x^2 \rangle_{\rm c} < 
\frac{\hbar^2}{2
m_e \varepsilon_g} .  \label{swm3} \end{equation} Below, we investigate the 
trends in
both members of this inequality for our test--case materials.

\section{Calculated localization tensors} \label{sect:calcu}

We have studied several tetrahedrally coordinated crystalline materials, 
from the group IV, III--V, and II--VI, having the diamond and zincblende
structure. The first-principle calculations have been performed within
density-functional theory in the local-density approximation, using
pseudopotentials \cite{pseudop} and plane waves.  We implement a trivial
extension of the formulas presented above, using a rectangular unit cell
instead of a simple cubic one: we thus describe the diamond and zincblende
structures by means of a tetragonal cell with a lattice constant $a$ in the
basal plane and $c = \sqrt{2} a$. There are four atoms per unit cell, whose
projections on the $c$ axis are equispaced; for the sake of consistency with
the formal results, we take $x$ along the $c$ axis and $yz$ in the basal
plane. We then use a BvK cell of sides $M_x c$, $M_y a$, and $M_z a$,
corresponding to a mesh in reciprocal space with $M_x,M_y,M_z$ points: this
allows an easier control of convergence.  

We start evaluating at the mesh points the Hermitian matrices \begin{equation} 
A_s =
S^\dagger ({\bf q}_s, {\bf q}_s\!+\!\Delta {\bf q}^{(x)}) S({\bf q}_s, {\bf
q}_s\!+\!\Delta {\bf q}^{(x)}) ; \label{hermi} \end{equation} then 
Eqs.~(\ref{facto3}) and (\ref{mdiago})
are written as \begin{equation} \langle x^2 \rangle_{\rm c} = \frac{1}{M_y M_z}
\sum_{s_2=1}^{M_y} \sum_{s_3=1}^{M_z} \left( - \frac{c^2 M_x}{4 \pi^2 n_b}
\sum_{s_1=1}^{M_x} \mbox{ ln det } A_s \right) . \label{noncubic} 
\end{equation}  In
general convergence is fast in $M_y$, $M_z$, and slower in $M_x$. The
expression in parenthesis in Eq.~(\ref{noncubic}) is precisely the 
one--dimensional
expression discussed in detail by RS, and the three dimensional one simply
obtains from it as an average in the $(q_y,q_z)$ plane.

\begin{figure}[tbp]
\begin{center}
\epsfig{figure=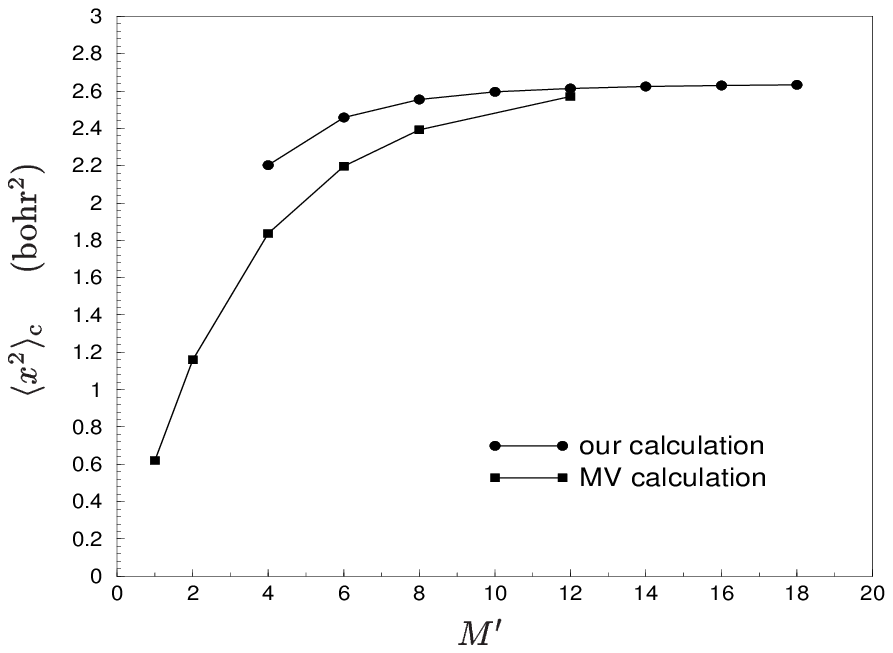,width=8.4cm}
\end{center}
\caption{Convergence of the squared localization length with the size of the
sampling grid, for the case of GaAs. 
We compare our discretized formula with the one used by MV,
using a genuinely cubic grid: the label $M'$ means 
$M' \times M' \times M'$ within MV notations.  }
\label{fig:MV} \end{figure}

First we show in Fig.~\ref{fig:MV} the convergence of our expressions over a
genuinely cubic grid, which coincides with the one used by MV in their
evaluation of the quantity $\Omega_{\rm I} = 3 n_b \langle x^2 \rangle_{\rm
c}$. They use a different discretization of the same ${\bf k}$ space
integral: both calculations converge to the same localization tensor,
although our discretization, based on Eqs.~(\ref{facto3}) and (\ref{mdiago}), 
converges faster.
All the following results have been obtained using noncubic grids, as in
Eq.~(\ref{noncubic}), in order to achieve faster convergence.

\begin{figure}[tbp] 
\begin{center}
\epsfig{figure=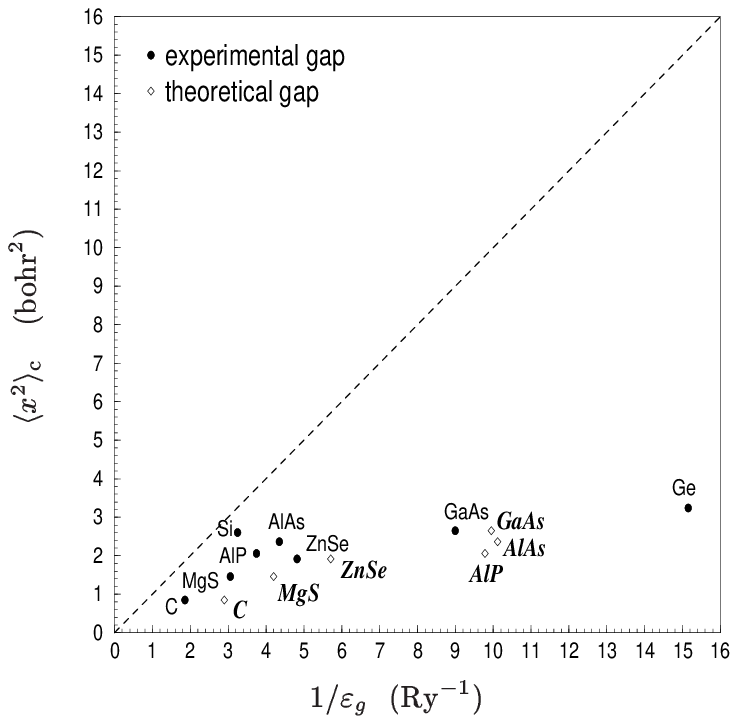,width=8.4cm}
\end{center}
\caption{Squared localization length vs. the inverse direct gap
(theoretical and experimental), for several elemental and binary
semiconductors. The inequality of Eq.~(\protect\ref{swm3}) is strongly 
verified. The points corresponding to Si and Ge with the theoretical gaps
are out of scale.}
\label{fig:binary} \end{figure}

We have systematically calculated well converged localization tensors for
several elemental and binary semiconductors. In Fig.~\ref{fig:binary} we plot
the localization tensors versus the right--hand member of the inequality in
Eq.~(\ref{swm3}), where for the gap $\varepsilon_g$ we have used both (i) the
theoretical and (ii) the experimental values. In case (i) the inequality owes
to an exact sum rule and {\it must} be satisfied:  we are therefore
checking the internal consistency of the calculations. Also, it may be
noticed that the inequality is very strongly verified. In case (ii) there is
no a--priori guarantee that the inequality is verified, particularly given
the fact that the experimental gap is systematically larger than the KS one.
Nonetheless the localization tensor is obtained here as a pure ground state
property, and it is well known that density-functional theory in the
local-density approximation provides a good representation of the ground
state, though not of the excitations.\cite{GW} It is therefore interesting to
verify that even for case (ii) the inequality in Eq.~(\ref{swm3}) is strongly
verified for all the materials considered.

\begin{figure}[tbp] 
\begin{center}
\epsfig{figure=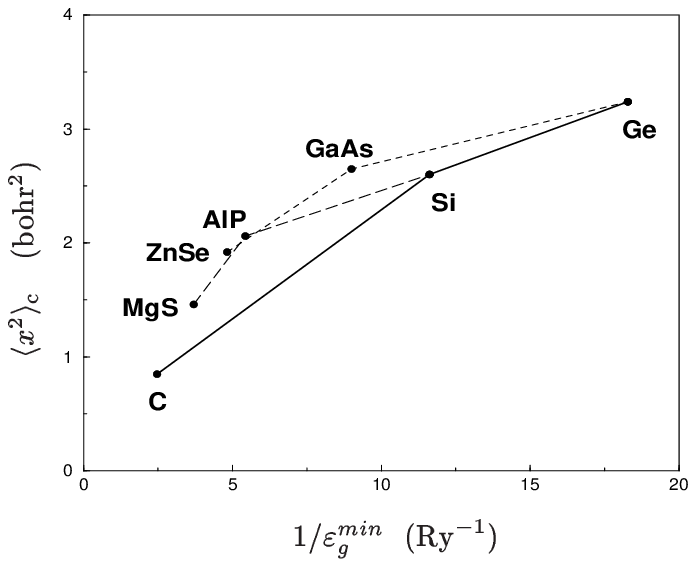,width=8.4cm}
\end{center}
\caption{Trends of the squared localization length vs. the inverse 
experimental minimum gap. The lines connect the isoelectronic
series MgS--AlP--Si and ZnSe--GaAs--Ge, and the isovalent one C--Si--Ge.}
\label{fig:families} \end{figure}

The localization tensor ranges roughly between 1 and 3 bohr$^2$ for all the
materials considered, diamond being the most localized and germanium the most
delocalized. The trend is qualitatively expected, in agreement with SWM's
statement that ``the larger the gap, the more localized the electrons are''.
However this is a trend more than a strict rule, and indeed a few materials
show irregularities. Better trends are obtained when comparing families of
materials: either isoelectronic series or isovalent series. In order to
enhance such regularities, we have heuristically tried a few different laws. 
In Fig.~\ref{fig:families} we plot our localization tensor versus
$1/\varepsilon_g$, using  the minimum gaps instead of the direct ones:
here monotonical trends are very perspicuous.

\section{Maximally localized hermaphrodite orbitals} \label{sect:herma}

We actually perform localizing transformations on the Bloch orbitals. At
variance with the most standard approach, we focus on orbitals which are
localized in one direction only, say $x$, while they are completely
delocalized in the $yz$ directions. By analogy with the standard theory of
Wannier functions,\cite{Blount} one obtains such orbitals by integration of
the Bloch ones over one component only of the Bloch vector. The resulting
orbitals are Wannier--like in one direction and Bloch--like in the other two:
they can be therefore called ``hermaphrodite orbitals''. Because of the same
reasons as for ordinary Wannier functions, such hermaphrodite orbitals are
nonunique: we focus here on those hermaphrodite orbitals which are optimally
localized in the $x$ direction.  It has been shown above that, in the
thermodynamic limit, these orbitals are eigenfunctions of the operator $\Xi$,
Eq.~(\ref{proj2}), and their centers are the corresponding eigenvalues. It is
expedient to consider the modified operator \begin{equation} \tilde{\Xi}({\bf
r, r'}) = \int_{\mbox{\scriptsize all space}} \!\!\!\!\!\!\!\!\!\!\!\! d{\bf
r''} \; P({\bf r, r''}) {\rm e}^{i \frac{2\pi}{M_x c} x''} P({\bf r'', r'})
\label{proj3} , \end{equation} which to leading order in $1/M_x$ has the same
eigenfunctions as $\Xi$, and simply related eigenvalues.

When considering a finite sample with BvK boundary conditions---or
equivalently a discrete grid in the reciprocal unit cell---the operator $\Xi$
as in Eq.~(\ref{proj2}) is useless because the operator $x$ therein becomes 
ill defined. Instead the operator $\tilde{\Xi}$ is well defined, provided the
value of $M_x$ is consistent with the choice of the grid. The integral in
Eq.~(\ref{proj3}) is now performed over the BvK cell and {\it not} over all 
space;
the projector projects over the finite occupied manifold, having dimension
$n_b M_x M_y M_z$. Choosing the Bloch functions as the basis in the occupied
manifold, the matrix elements of $\tilde{\Xi}$ are nothing else than the
matrix ${\cal S}$ defined in Eq.~(\ref{overlap}). Therefore in the discrete 
case the hermaphrodite orbitals which achieve optimal localization in the $x$
direction are simply obtained by diagonalizing the matrix ${\cal S}$. Since,
as already observed, the matrix is already diagonal in $s_2$ and $s_3$, the
problem is reduced to $M_y M_z$ independent diagonalizations of submatrices
of size $n_b M_x$. We characterize our orbitals as $w_{j,s_2,s_3}$, where
$(s_2,s_3)$ is a two--dimensional Bloch--index and $j$ is a one--dimensional
Wannier--like index, with $1 \leq j \leq n_b M_x$.

We are going to verify that these orbitals indeed minimize the average
quadratic spread in one--dimension. If we define \begin{equation} 
z_{j,s_2,s_3} =
\int_{\mbox{\scriptsize BvK cell}} \!\!\!\!\!\!\!\! d {\bf r} \;
|w_{j,s_2,s_3}({\bf r})|^2 {\rm e}^{i\frac{2\pi}{M_x c} x} , \label{dim1} 
\end{equation}
then according to RS  the quadratic spread of one given hermaphrodite orbital
is: \begin{equation} \langle w_{j,s_2,s_3} | x^2 | w_{j,s_2,s_3} 
\rangle_{\rm c} = -
\frac{c^2 M_x^2}{4 \pi^2} \ln | z_{j,s_2,s_3} |^2 . \label{dim2} \end{equation}
 
Taking
now the average over all orbitals, and calling this quantity
$\lambda^2_{xx}$, we get \begin{equation} \lambda^2_{xx} = \frac{1}{M_y M_z}
\sum_{s_2=1}^{M_y} \sum_{s_3=1}^{M_z} \left( - \frac{c^2 M_x}{4 \pi^2 n_b}
\sum_{j=1}^{n_b M_x} \ln | z_{j,s_2,s_3} |^2 \right) . \label{lambda}
\label{dim3} \end{equation} We then notice that $w_{j,s_2,s_3}$ are the 
eigenvectors of
$\tilde{\Xi}$, hence the expectation values $z_{j,s_2,s_3}$ are the
corresponding eigenvalues. Since the product of the eigenvalues equals the
determinant, standard manipulations prove that the average spread
$\lambda^2_{xx}$ equals indeed the lower bound $\langle x^2 \rangle_{\rm c}$
as given in Eq.~(\ref{noncubic}).

There is a subtlety about the diagonalization of the submatrices of ${\cal
S}$, which are the projection over a certain finite dimensional manifold of
the operator ${\rm e}^{i\frac{2\pi}{M_x c} x}$. Although the operator is
unitary, its projection is not a unitary matrix, hence the eigenvectors are
not exactly orthogonal, as instead honest localized orbitals must be: this is
not a serious problem. In fact the larger is $M_x$, the closer to unitarity
the matrix becomes: we know that the modulus of its determinant differs by
one for a term of the order $1/M_x$, hence the modulus of each eigenvalue
differs by one for a term of the order $1/M_x^2$. We recover exact
orthonormality in the thermodynamic limit; in our calculations already at
$M_x \simeq 20$ deviations from orthogonality are hardly noticeable.

\begin{figure}[tbp]
\begin{center}
\epsfig{figure=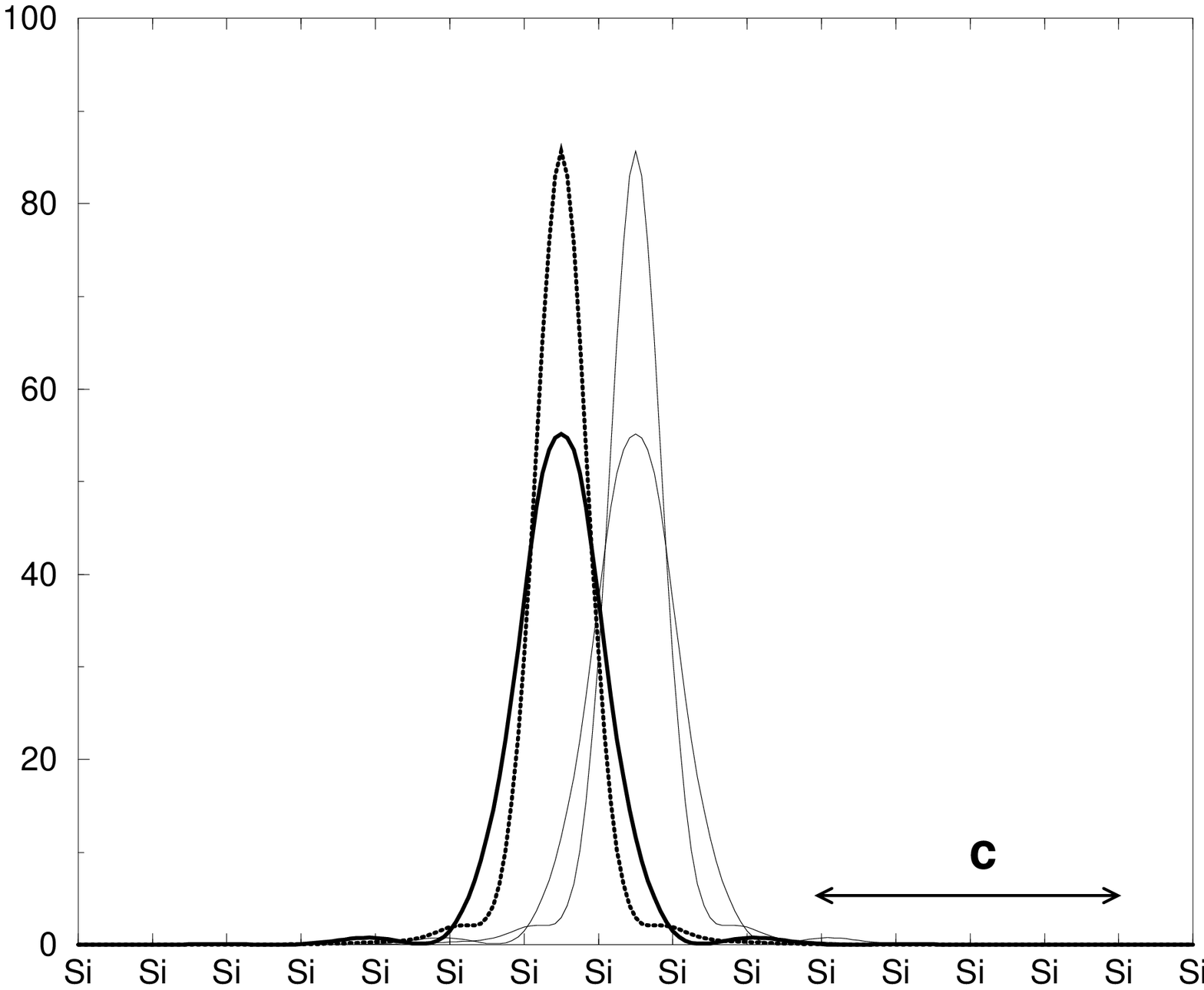,width=8.4cm}
\end{center}
\caption{\label{fig:siherma} Hermaphrodite orbitals for Si. The quantity
displayed is $n_{\rm loc}(x)$, defined as the $yz$ average of the square
modulus of the orbital $w_{j,s_2,s_3}$, for $s_2=s_3=0$, and for the four
$j$ values localizing within the same cell.  }

\end{figure}

\begin{figure}[tbp]
\begin{center}
\epsfig{figure=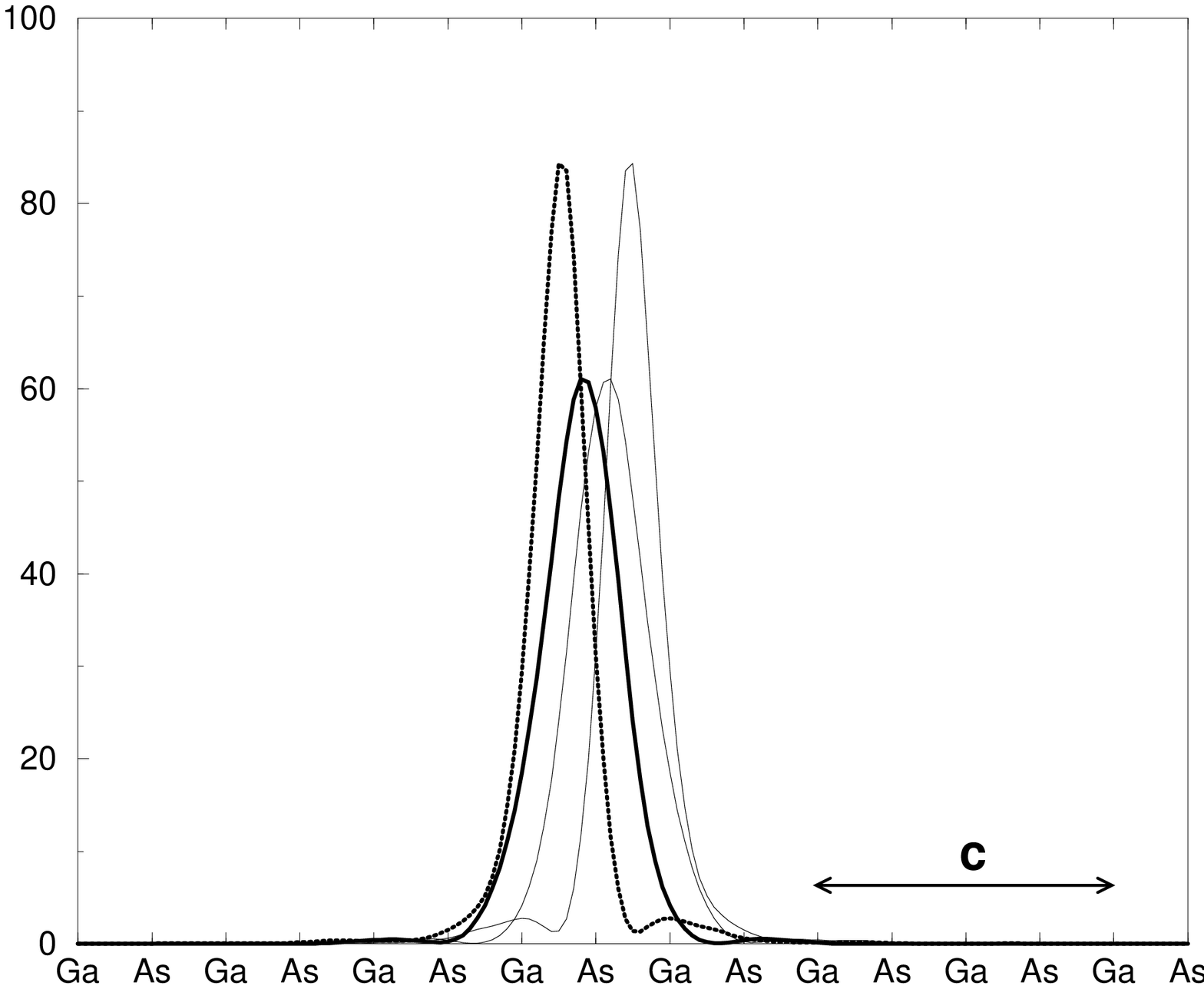,width=8.4cm}
\end{center}
\caption{\label{fig:gaasherma} Hermaphrodite orbitals for GaAs. The quantity
displayed is $n_{\rm loc}(x)$, defined as the $yz$ average of the square
modulus of the orbital $w_{j,s_2,s_3}$, for $s_2=s_3=0$, and for the four
$j$ values localizing within the same cell.  }
\end{figure}

We have calculated the orbitals $w_{j,s_2,s_3}$ for several crystalline
semiconductors: to the purpose of display, we call the $yz$ average of
$|w_{j,s_2,s_3}|^2$ as $n_{\rm loc}(x)$, where the indices remain implicit. 
At fixed $(s_2,s_3)$ we have, given our double cell, $8 M_x$ orbitals
centered on a BvK period of length $M_x c$. There are however at most four
different shapes, and one obtains all the functions upon translations in the
$x$ direction (by multiples of $c/2$) of the four basic ones: this is not
surprising, since the genuine unit cell is one half of our computational one.
We find that the different shapes are actually always four, with the only
exception of an elemental semiconductor at $s_2=s_3=0$. In this very special
case the different shapes are only two, the orbitals are centered at the bond
center, and their densities are centrosymmetric: the corresponding functions
$n_{\rm loc}(x)$ are shown in Fig.~\ref{fig:siherma} for the case of Si. The
most general case is exemplified by Fig.~~\ref{fig:gaasherma}: it shows the
four different $n_{\rm loc}(x)$ for the case of GaAs, again at $s_2=s_3=0$. 
None of these four $w$ orbitals is therefore centered at a symmetry site, and
none is centrosymmetrical, although they are obviously symmetrically related
to each other. About the actual value of the quadratic spread (in the $x$
direction) of each of the $w$'s, we have found as a general feature that the
least localized ones are those for $s_2=s_3=0$, {\it i.e.} at the $\Gamma$
point in the two--dimensional reciprocal space.

We now address the long standing issue of exponential localization. Exact
general results only exist for the genuine one--dimensional case, where W. 
Kohn has proved long ago\cite{Kohn59} that the Wannier functions which
minimize the quadratic spread ({\it i.e.} are optimally localized) have an
asymptotic exponential behavior. After the present work was completed, we
became aware of Ref.~\onlinecite{He01}, where the asymptotic exponential is
shown to have a power--law prefactor.  In three dimensions the problem is
unsolved, with the exception of some very special cases.\cite{desCloizeaux64}
It has been conjectured by MV that their optimally localized Wannier
functions enjoy three--dimensional exponential localization: an analytical
proof looks very hard. Our hermaphrodite orbitals $w$ are optimally localized
in the $x$ direction, and therefore in a sense they have one--dimensional
character: nonetheless, they are have a genuine dependence on all three
coordinates.  Therefore an analytic proof along the lines of
Refs.~\onlinecite{He01} and \onlinecite{Kohn59} is not easily extended to our
case. Instead, it is simple to use the same arguments as given in
Ref.~\onlinecite{Niu91} in order to prove that our $w$ orbitals decay in the
$x$ direction faster that any inverse power of $x$.

\begin{figure}[tbp]
\begin{center}
\epsfig{figure=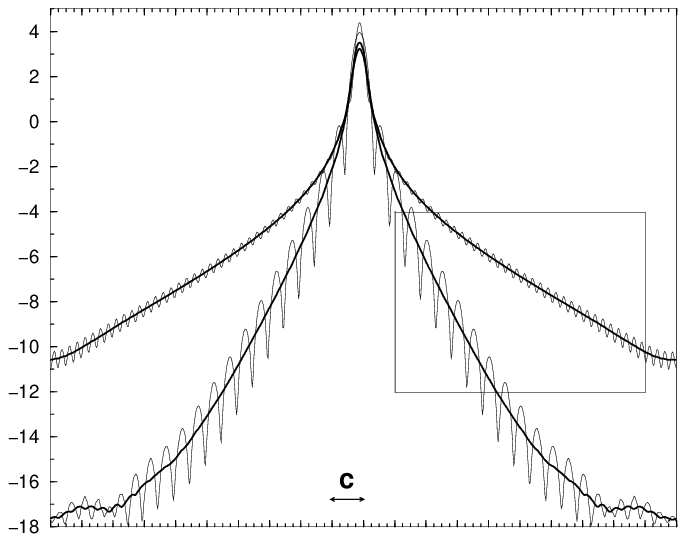,width=8.4cm}
\epsfig{figure=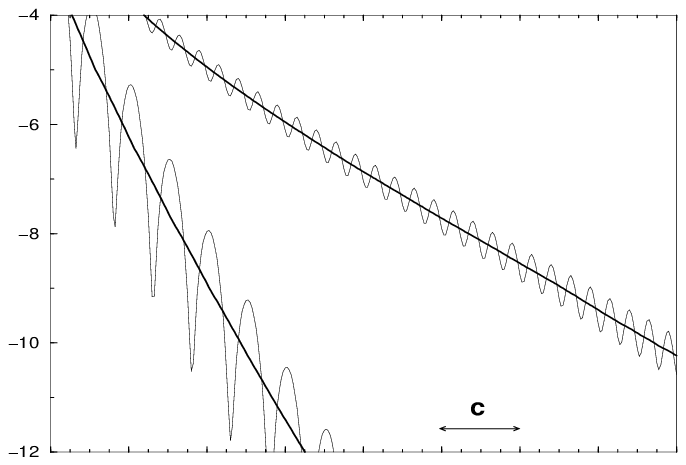,width=8.4cm}
\end{center}
\caption{\label{fig:asymp} Exponential decay of two 
hermaphrodite orbitals for Ge. Thin lines correspond to the logarithm of
two different $n_{\rm loc}(x)$, defined as the $yz$ average of the square
modulus of the orbital $w_{j,s_2,s_3}$, for the same $j$ and 
two different points of the two-dimensional mesh($s_2$,$s_3$). 
The one with the slowest decay 
corresponds to  $s_2=s_3=0$. Thick lines are the macroscopic average
(see text) of $\ln n_{\rm loc}(x)$. The lower panel is a magnification of
the region indicated in the upper panel in order to better appreciate 
the linear behavior of $\ln n_{\rm loc}(x)$.}
\end{figure}

\begin{figure}[tbp] \begin{center} \epsfig{figure=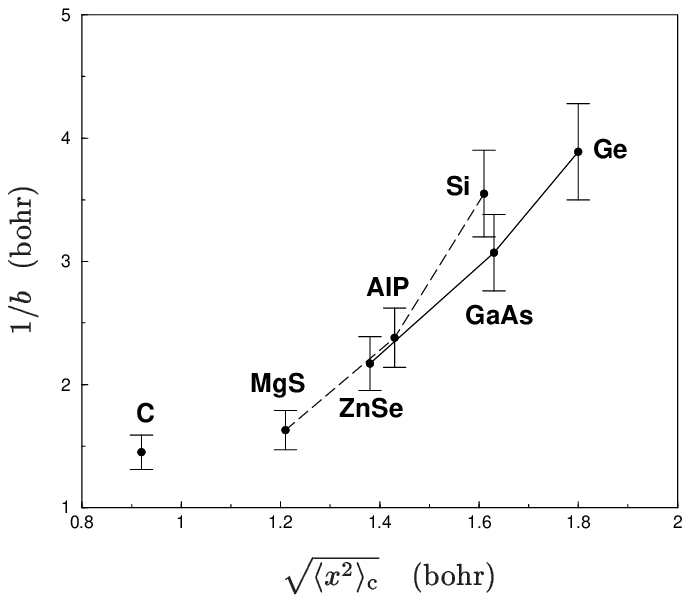,width=8.4cm}
\end{center} \caption{\label{fig:trends2} Exponential decay length, averaged
over the two-dimensional mesh ($s_2,s_3$), vs. our localization length
(square root of the second cumulant moment). The straight-line segments are
only a guide to the eye linking compounds of the same isoelectronic series.
The vertical bars are an estimate of the accuracy of the interpolation scheme
used to extract the $b$ value from the asymptotic macroscopic average of
$n_{\rm loc}(x)$. }
\end{figure}

Our very elongated BvK cells allow us to study the asymptotic behaviour
heuristically on our calculated $w$'s. The nearest periodic replica of a
given $w$ orbital is centered at a distance of $M_x c$ from it: therefore the
interesting ``asymptotic region'' is accessible up to a distance somewhat
smaller than $M_x c/2$, as it clearly appears from Fig.~\ref{fig:asymp}.  The
quantity of choice in order to ``blow up'' the exponential behavior is
obviously the logarithm of $n_{\rm loc}(x)$, which we plot in
Fig.~\ref{fig:asymp} (thin solid lines) for the case of Ge and for two
different $(s_2,s_3)$.  It is seen that there is a wide region where the
plots have an overall linear behavior, with superimposed oscillations having
the crystal periodicity along the $x$ direction ($c/2$ in the present case). 
The slopes at different $(s_2,s_3)$ are very different, though; the $n_{\rm
loc}(x)$ with the slowest decay corresponds to $s_2=s_3=0$ and therefore to
the least localized, as we previously observed.  Next, we filter the
disturbing periodic oscillations using our favorite tool of the macroscopic
average.\cite{macro} We tried both ways: filtering $n_{\rm loc}(x)$ first
and then taking the logarithm, or filtering $\ln n_{\rm loc}(x)$: the latter
turns out to work best. The macroscopic filtering is also shown in
Fig.~\ref{fig:asymp} (thick solid lines): it is easily realized, expecially
looking at the magnified plot in the lower panel, that there is a sizeable
region, spanning several cells, where the plotted function looks accurately
linear with $x$, hence $n_{\rm loc}(x) \propto \exp(\pm bx)$.  We therefore
demonstrate ``experimentally'' the exponential localization of our $w$
orbitals. After we became aware of Ref.~\onlinecite{He01}, we checked that
the power--law prefactor suggested therein does not improve the quality of our
fits.  It is hard to assess whether this is due to a basic difference between
our case and a genuine one--dimensional one, or to the limited resolution
achievable in our selfconsistent three-dimensional finite--size calculation.
Finally, in Fig.~\ref{fig:trends2}  we display some correlations between the
localization length and the exponential decay length $1/b$ averaged over the
two-dimensional mesh ($s_2,s_3$).

\section{Conclusions}   \label{sect:conclu}

In the present work we provide the three--dimensional formulation of the RS
theory of electron localization,\cite{rap107} specializing it to the case of
independent KS electrons; we discuss some of its relationships to the MV and
SWM papers,\cite{Marzari97,Souza00} and also (in the Appendix) how it relates
to Boys theory of localization in molecules.\cite{Boys60} We then implement
the theory to several materials in the class of tetrahedrally coordinated
semiconductors.  Among the results, we find that in general the
calculated localization length is a monotonical function of the gap, although
a few materials show irregularities. The trend is more regular within a 
given family (isoelectronic or isovalent). 
Finally, we heuristically show that the orbitals which are optimally
localized in a given direction (``hermaphrodite orbitals'') show exponential
localization.

\section*{Acknowledgments}

R.R. acknowledges partial support by ONR grant
N00014-96-1-0689.

\appendix

\section*{Relationship to Boys localization in molecules}

We abandon here the BvK boundary conditions used throughout this
work, and we consider an  $N$--electron system which is bounded in space. 
Both the orbitals and the many--body wavefunction $\Psi$ are therefore
exponentially vanishing at large distances. Supposing that $N$ is even and
the state is a singlet, for independent particles the wavefunction is the
Slater determinant: \begin{equation} \Psi = \frac{1}{N!} |\, \varphi_{1}
\overline{\varphi}_{1} \varphi_{2} \overline{\varphi}_{2} \dots \varphi_{N/2}
\overline{\varphi}_{N/2}\, | .  \label{slater1} \end{equation} The orbitals 
enjoy no
specific symmetry. Any unitary transformation of the orbitals produces the
same many--body ground state (modulo an overall phase): a specific choice of
the orbitals will be referred to as ``choice of the gauge'' in the following.
Obviously all ground state properties are gauge--invariant. The
density matrix is twice the projector over the occupied orbitals: 
\begin{equation}
\rho({\bf r, r'}) = 2 P({\bf r, r'}) = 2 \sum_{i=1}^{N/2} \varphi_{i}({\bf
r}) \varphi_{i}^*({\bf r'}) \label{projm} . \end{equation}

We are interested in exploiting the gauge freedom in order to express the
ground state in terms of localized orbitals.\cite{Pipek89} The standard Boys
localization\cite{Boys60} in molecules consists in minimizing spherical
second moments, in perfect analogy with MV, which can be regarded as the
solid--state analogue of Boys localization. Here instead we are mostly
interested in localizing in one given direction, say $x$. 

For any given choice of the single-particle orbitals $\varphi_i$, the average
quadratic spread in the $x$ direction is by definition: \begin{equation} 
{\lambda}^2_{xx} =
\frac{2}{N} \sum_{i=1}^{N/2} ( \langle \varphi_i | x^2 | \varphi_i \rangle -
\langle \varphi_i | x | \varphi_i \rangle^2 ) .  \end{equation} We recast this
identically as: \begin{eqnarray} {\lambda}^2_{xx} = \frac{2}{N} \sum_i
\langle \varphi_i | \, x \, ( \, 1 - \sum_j | \varphi_j \rangle \langle
\varphi_j | \, ) \, x \, | \varphi_i \rangle \nonumber \\ + \frac{2}{N}
\sum_{i \neq j} | \langle \varphi_i | x | \varphi_j \rangle |^2 .  \qquad
\qquad \qquad \label{tildel} \end{eqnarray} The first term in 
Eq.~(\ref{tildel}) is
gauge invariant, since we can identically write: \begin{equation} 
{\lambda}^2_{xx} =
\frac{2}{N} \mbox{Tr } x P x (1 - P) + \frac{2}{N} \sum_{i \neq j} | \langle
\varphi_i | x | \varphi_j \rangle |^2 , \label{tildell} \end{equation} where 
``Tr''
indicates the trace on the electronic coordinate. The gauge--invariant term
in Eq.~(\ref{tildell}) can be regarded as the $xx$ element of a more general
tensor, which turns out to be the molecular analogue of our localization
tensor. We use the same notation for molecules and for crystals: 
\begin{equation} \langle
r_\alpha r_\beta \rangle_{\rm c} = \frac{2}{N} \mbox{Tr } r_\alpha P r_\beta
(1 - P) . \label{trace} \end{equation} 

If we look for the orbitals which minimize the average spread in the $x$
direction, the solution, after Eq.~(\ref{tildell}), is provided by those 
orbitals
which diagonalize the position operator $x$, projected over the occupied
manifold.  Obviously, a set of orthonormal orbitals which diagonalize $PxP$
can always be found, since $PxP$ is a Hermitian operator.  The quadratic
spread of these orbitals equals $\langle x^2 \rangle_{\rm c}$, the
gauge--invariant part in Eq.~(\ref{tildell}). If we are interested instead in
minimizing the spherical second moment, in general we {\it cannot}
diagonalize simultaneously $PxP$, $PyP$, and $PzP$.  Therefore the spherical
spread will be in general {\it strictly larger} than the Cartesian trace of
the localization tensor. This is a key feature in the work of
Boys,\cite{Boys60} and MV as well.

We have defined the localization tensor in Eq.~(\ref{trace}). With an obvious
generalization of the previous arguments, this tensor provides in general the
maximum localizability {\it in any given direction}. An equivalent expression
for the localization tensor is: \begin{equation} \langle r_\alpha r_\beta 
\rangle_{\rm c} =
\frac{1}{N} \int d {\bf r} \int d {\bf r}' \; ({\bf r - r'})_\alpha  ({\bf r
- r'})_\beta \, |P({\bf r}, {\bf r}')|^2 \label{short1} , \end{equation} 
which has the
meaning of the second moment of the (squared) density matrix in the
coordinate ${\bf r - r'}$.  

At this point, we may think of a crystalline solid as of a very large
``molecule'', or a cluster, and take the thermodynamic limit. Since bulk
properties must be independent of the choice of boundary conditions (either
BvK or ``free''), the density matrix and the localization tensor must be the
same as the one previously found in this work. And indeed, a glance to
Eq.~(\ref{mproj}) shows that it coincides with the thermodynamic limit of
Eq.~(\ref{short1}) in the insulating case. As for the metallic case, our 
previous
findings bear an important message concerning Boys localization. For a
cluster of finite size, no matter how large, one can doubtless build
localized Boys orbitals. But our results prove that, in the large $N$ limit,
the quadratic spread of these Boys orbitals diverges whenever the cluster is
metallic.

\begin{thebibliography}{10}
\bibitem{GW} For instance, when implementing the GW method in semiconductors,
the occupied orbitals are routinely approximated with the KS ones. The
calculated excitation properties are much closer to the experimental
ones, though.
\end{thebibliography}
\end{document}